\documentclass[aps,floatfix,twocolumn,showpacs]{revtex4}
\usepackage{natbib}
\usepackage{dcolumn}
\usepackage{graphicx}

\newcommand{\cals}{\zeta_s}
\newcommand{\calr}{\zeta_r}
\newcommand{\calx}{\zeta_{s,r}}

\newcommand{\WMAP}{{\slshape WMAP~}}

\newcommand{\apjl}{Astrophys.~J.~Lett.}
\newcommand{\mnras}{Mon.~Not.~R.~Astron.~Soc.}
\newcommand{\aap}{Astron.~Astrophys.}
\newcommand{\aj}{Astron.~J.}
\newcommand{\physrep}{Phys.~Rep.}
\newcommand{\araa}{Ann.~Rev.~Astron.~Astrophys.}
\newcommand{\apjs}{Astrophys.~J.~Supp.}

{ }

\begin{document}

\title{Spectroscopic source redshifts and parameter constraints from 
  weak lensing and CMB}

\author{ Mustapha Ishak$^{1}$ }
\author{ Christopher M. Hirata$^{2}$ }
\affiliation{
$^1$Department of Astrophysical Sciences, Princeton University,
  Princeton, NJ 08544, USA \\
$^2$Department of Physics, Princeton University,
  Princeton, NJ 08544, USA} 

\date{\today}

\begin{abstract}
Weak lensing is a potentially robust and model-independent cosmological
probe, but its accuracy is dependent on knowledge of the redshift
distribution of the source galaxies used.  The most robust way to
determine the redshift distribution is via spectroscopy of a subsample of
the source galaxies.  We forecast constraints from combining CMB
anisotropies with cosmic shear using a spectroscopically determined
redshift distribution, varying the number of spectra $N_{spec}$ obtained
from 64 to $\infty$.  The source redshift distribution is expanded in a
Fourier series, and the amplitudes of each mode are considered as
parameters to be constrained via both the spectroscopic and weak lensing
data. We assume independent source redshifts, and consider in what circumstances 
this is a good approximation (the sources are clustered and for narrow spectroscopic 
surveys with many objects this results in the redshifts being correlated).
It is found that for the surveys considered and for a prior 
of  0.04 on the calibration parameters, the addition of redshift 
information make significant improvements on the constraints on
the cosmological parameters; however, beyond $N_{spec}\sim$few$\times 10^3$
the addition of further spectra will make only a very
small improvement to the cosmological parameters.  We find that a better
calibration makes large $N_{spec}$ more useful.  Using an eigenvector analysis, 
we find that the improvement continues with even 
higher $N_{spec}$, but not in directions that dominate the uncertainties 
on the standard cosmological parameters.
\end{abstract}
\maketitle
%
\section{Introduction}
\label{sec:intro}
Weak lensing (WL) is a promising tool for an era of precision cosmology 
(for reviews, see
\cite{2003ARA&A..41..645R,2003astro.ph..5089V,2001PhR...340..291B,
1999ARA&A..37..127M} and
references therein).  Many recent studies showed the potential of this
established technique in constraining various cosmological parameters
\cite{1999ApJ...514L..65H, 2002PhRvD..65b3003H, 2002PhRvD..65f3001H,
2003PhRvL..91d1301A, 2003astro.ph..6033B, 2004MNRAS.348..897T,
2003MNRAS.343.1327H, 2003PhRvL..91n1302J, 2004ApJ...601L...1T,
2004ApJ...600...17B, 2004PhRvD..69h3514I, 2004A&A...417..873S}.
Using current data,
Refs.~\cite{2003PhRvL..90v1303C, 2002A&A...393..369V, 2003PhRvD..68l3001W,
2003AJ....125.1014J, 2004astro.ph..4195M}
showed that WL can provide constraints that are competitive
with other cosmological probes.  Many larger and more ambitious surveys 
are ongoing, planned or proposed.  These include the Deep Lens Survey 
\cite{2002SPIE.4836...73W}; the NOAO Deep Survey; the
CFHT Legacy Survey \cite{2001misk.conf..540M}; Pan-STARRS; {\slshape SNAP}
\cite{2004APh....20..377R, 2004AJ....127.3089M, 2004AJ....127.3102R}; and
LSST \cite{2002SPIE.4836...10T}.
A major challenge for the future of WL  studies is to have a very
tight control of systematic errors.  
These include incomplete knowledge of the source redshift
distribution \cite{2000Natur.405..143W,2003ARA&A..41..645R}.  Recent
cosmic shear studies have obtained their redshift distributions from joint
redshift-magnitude distributions measured in independent surveys
\cite{2003AJ....125.1014J}; from photometric redshifts (``photo-$z$'s'')
\cite{2003MNRAS.341..100B}; or some combination.  The former approach
suffers from the difficulty that the selection function (and relative
weighting) of galaxies may be different in the lensing and redshift
surveys, while the latter approach suffers from possible photo-$z$ errors
that are difficult to constrain in the absence of spectroscopic
confirmation, especially if only a small number of colors are measured.  
Several studies have marginalized over the source redshift distribution.
Ultimately the most robust and model-independent determination of the
source redshift distribution would be via spectroscopy of a randomly
chosen sub-sample of the source catalog.  However, spectroscopy of faint
galaxies can be very time-consuming even with today's large-aperture
telescopes. In this paper, we forecast Fisher-matrix constraints
from cosmic shear using a spectroscopically determined redshift
distribution, varying the number of spectra $N_{spec}$ obtained.  The
source redshift distribution is expanded in a Fourier series, and the
amplitudes of each mode are considered as parameters to be constrained via
both the spectroscopic and WL data. We address the effect of galaxy clustering
on our analysis and discuss the effect of a better calibration.

\section{Model parameters}
The following basic parameter set for WL is considered:
$\Omega_{m}h^2$, the physical matter density; $\Omega_\Lambda$ and $w$,
respectively the fraction of the critical density in a dark energy
component and its equation of state;  $n_s(k_0=0.05h/{\rm Mpc})$
and $\alpha_s$, the spectral index and running of the primordial scalar
power spectrum at $k_0$; $\sigma_8^{\rm lin}$, the amplitude of linear
fluctuations; 
 $\{q_{1}...q_{j_{max}}\}$, the leading
coefficients of a Fourier expansion of the source galaxy redshift
distribution (see Eq. \ref{eq:z_dist_expansion} in the see next section); 
in the case of two bin tomography we use $\{q_{jA},q_{jB}\}$ with $j=1..j_{\rm max}$. 
We also include $\cals$ and $\calr$ as defined in \cite{2004PhRvD..69h3514I} 
to parametrize the shear calibration bias 
\cite{2001A&A...366..717E, 2001MNRAS.325.1065B, 2003MNRAS.343..459H,
2002AJ....123..583B, 2003astro.ph..5089V, 2000ApJ...537..555K}, in which
the gravitational shear is systematically over- or under-estimated by
a multiplicative factor, {\it i.e.} 
$\hat P_\kappa(\ell) = (1+\cals) P_\kappa(\ell)$,  
where $P_\kappa(\ell)$ is the convergence power spectrum obtained in the absence
of calibration errors. $\cals$ refers to the calibration
error of the power spectrum, which is twice the calibration error of
the amplitude as the power spectrum is proportional to
amplitude squared.
When we consider tomography, we must also consider
the relative calibration $\calr$ between the two redshift
bins.  This error affects the measured power spectrum $\tilde
P_\kappa(\ell)$ in accordance with:
\begin{eqnarray}
\tilde P_\kappa^{AA} = && \!\!\!\! (1+f_B \calr) \hat P_\kappa^{AA}(\ell),
\nonumber \\
\tilde P_\kappa^{AB} = && \!\!\!\! (1+{f_B-f_A\over 2} \calr) \hat P_\kappa^{AB}(\ell),
\nonumber \\
\tilde P_\kappa^{BB} = && \!\!\!\! (1-f_A \calr) \hat P_\kappa^{BB}(\ell),
\label{eq:calr}
\end{eqnarray}
where $f_A$  and $f_B$ are the fraction of the source
galaxies in bin A and B respectively.

In order to combine this with information from the CMB we include $\Omega_{b}h^2$, the physical baryon density; $\tau$, the optical 
depth to reionization; and $T/S$, the tensor-to-scalar fluctuation ratio.  We assume a spatially flat Universe with 
$\Omega_{m}+\Omega_{\Lambda}=1$, thereby fixing $\Omega_m$ and $H_0$ as functions of our basic parameters, and we do not include 
massive neutrinos, or primordial isocurvature perturbations. We use as fiducial model ({\it e.g.} Ref. \cite{2003ApJS..148..175S}, 
with $w$ and $T/S$ added): $\Omega_bh^2=0.0224$, $\Omega_m h^2=0.135$, $\Omega_{\Lambda}=0.73$, $w=-1.0$, $n_s=0.93$, $\alpha_s=0.0$, 
$\sigma_{8}=0.84$, $\tau=0.17$, $T/S=0.2$, $\cals=0.0$, $\calr=0.0$, and $q_{i},q_{jA},q_{jB}=0.0$.

We will consider surveys with $f_{sky}=0.01$ and $0.1$, and have number density-to-shape noise ratio $\bar 
n/\langle\gamma^2_{int}\rangle = 4.1\times 10^9$~sr$^{-1}$, corresponding to a number density of $\bar n=30$ galaxies/arcmin$^2$ and 
shape+measurement noise $\langle\gamma^2_{int}\rangle = (0.3)^2$.  (Note that this is the shape noise in the shear $\gamma$, rather 
than the ellipticity which is roughly $e\approx 2\gamma$ if isophotal or adaptive ellipticities are used \cite{2002AJ....123..583B, 
2004PhRvD..70f3510S}.)

\section{Parametrization of the redshift distribution}

\begin{figure}[ht]
\includegraphics[width=2.4in,angle=-90]{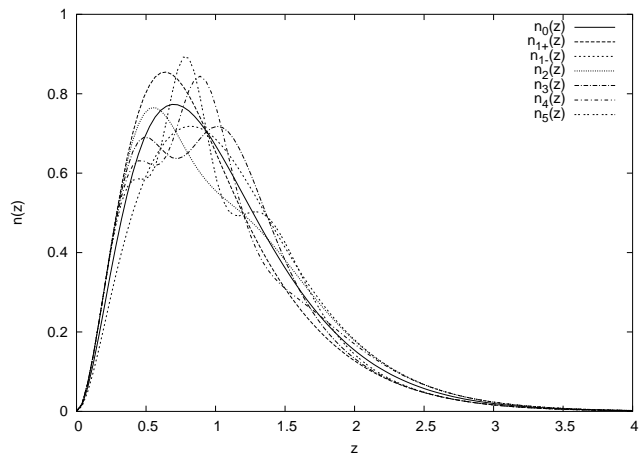}
 \caption{\label{fig:fig1} 
 Variations of the redshift distribution. The fiducial distribution 
 $n_0(z)$ (which peaks at $z_p=2z_0$) and the Fourier expansion terms 
are plotted. 
 The distributions $n_{1\pm }(z)$ represent $q_1$ varied
 by $\pm 25 \%$. We also plot $n_{2+}(z)$ to $n_{5+}(z)$.
 While we varied the $q_j's$ by $\pm 0.05$
 in the analysis, we plot the $25 \%$ variations just
 for the clarity of the plot.
}
\end{figure}

The source averaged distance ratio appears as
a weighting function in the convergence power spectrum
\cite{1992ApJ...388..272K,1997ApJ...484..560J,1998ApJ...498...26K} and
is given by
\begin{equation}
g(\chi) = \int_\chi^{\chi_H} n(\chi') {\sin_K(\chi'-\chi)\over
\sin_K(\chi')} d\chi',
\end{equation}
where $n(\chi(z))$ is the normalized source redshift distribution.
Deviations from the fiducial  distribution,  \cite{2000Natur.405..143W},
\begin{equation}
n_{0}(z)=\frac{z^2}{2 z_0^3}\, e^{-z/z_0} , 
\end{equation}
(which peaks at $z_p=2z_0=0.70$ and has $z_{med}\approx 0.94$ ) 
are parameterized via a Fourier series,
\begin{equation}
n(z)=n_{0}(z)\Big{[}1+\frac{1}{\sqrt{2}}\sum_{j=1}^\infty q_{j} \cos(j\pi 
P_{0}(z)) \Big{]},
\label{eq:z_dist_expansion}
\end{equation}
where $P_0$ is the cumulative redshift distribution in the fiducial model,
\begin{equation}
P_{0}(z)=1-\Big{(}1+\frac{z}{z_0}+\frac{z^{2}}{2z_{0}^{2}}\Big{)}e^{-z/z_0}
=\int_{0}^{z} n_{0}(z')dz'.
\label{eq:P_0}
\end{equation}
The $\{q_j\}$ are thus the coefficients in the Fourier expansion of
$n(z)/n_0(z)$ in the interval $0\le P_0<1$.  We have used the cumulative 
fiducial probability $P_0(z)$ instead of
$z$ as the independent variable because this will result in uncorrelated
constraints on the $q_j$ from spectroscopy; the
$j=0$ coefficient vanishes because of the normalization constraint $\int
n(z)dz=\int n_0(z)dz=1$.  The selection of the cosines is arbitrary and 
any complete set of orthogonal functions would work just as well.
The distribution is completely specified by the 
$\{q_j\}_{j=1}^\infty$, although in this paper we cut off the series at some $j_{\rm max}$
(we show results for $5$ and $100$).
For tomography, the normalized 
 distributions and the respective Fourier expansions are 
given as above, except for bin A we replace $n_{0}(z)$ with
\begin{eqnarray}
{n_{0}^{A}(z)
= \frac{n_{0}(z)}{1-5/e^{2}}} &
  {\;\;{\rm for}\;\;} {z_{p} \le 2 z_{0}}, \nonumber \\
{n_{0}^{A}(z)=0 } & \;\;{\rm for}\;\;
  {z_p > 2 z_0},
\label{eq:z_distA}
\end{eqnarray}
%
and for bin B, we replace $n_{0}(z)$ with
\begin{eqnarray}
{n_{0}^{B}(z)=0 } & \;\;{\rm for}\;\;
  {z_p \le 2 z_0},\nonumber \\
{n_{0}^{B}(z)= \frac{n_{0}(z)}{5/e^{2}}} &
  {\;\;{\rm for}\;\;} {z_{p} > 2 z_{0}}.
\label{eq:z_distB}
\end{eqnarray}
%
The cumulative probabilities for the two bins are
$P_{0}^{A}(z)=\int_{0}^{z} n_{0}^{A}(z')dz'$ and
$P_{0}^{B}(z)=\int_{0}^{z} n_{0}^{B}(z')dz'$.
We use the parameters $\{q_1..q_5,q_{1}^{A}..q_{5}^{A},q_{1}^{B}..q_{5}^{B}\}$ to vary the redshift 
distribution, see Fig. \ref{fig:fig1}. 
%
\section{Fisher-matrix analysis}
The statistical error on a given
parameter $p^{\alpha}$ is given by \begin{equation}
\sigma^2(p^{\alpha})\approx [({\bf F}_{CMB}+{\bf F}_{WL}+{\bf
F}_{spec}+\Pi)^{-1}]^{\alpha \alpha}, \label{eq:sigma2} \end{equation}
where $\Pi$ is the prior curvature matrix, and ${\bf F}_{CMB}$, ${\bf
F}_{WL}$ and ${\bf F}_{spec}$ are the Fisher matrices from CMB, WL, and
spectroscopy, respectively.  We use for ${\bf F}_{WL}$ the approach
described in Ref.~\cite{2004PhRvD..69h3514I}, with a cutoff at $\ell_{\rm
max}=3000$ since on smaller scales the trispectrum contribution to the WL
covariance (neglected in Eq.~\ref{eq:sigma2})  dominates
\cite{2000ApJ...530..547J, 2001ApJ...554...56C}. For CMB, we use the 4
year \WMAP parameter constraints including $TT$, $TE$, and $EE$ power
spectra, assuming $f_{sky}=0.768$ (the Kp0 mask of
Ref.~\cite{2003ApJS..148...97B}), temperature noise of $400$, $480$, and
$580\,\mu$K~arcmin in Q, V, and W bands respectively (the rms noise was
multiplied by $\sqrt{2}$ for polarization), and the beam transfer
functions of Ref.~\cite{2003ApJS..148...39P}. The spectroscopy Fisher
matrix is obtained as follows.  If a galaxy is chosen at random from the
lensing catalog, and is spectroscopically determined to have redshift $z$,
then the log-likelihood for the redshift distribution parameters $q_j$ is
\begin{equation} \ln{\cal L}(q_j) = \ln n_0(z) + \ln \Big[ 1 +
{1\over\sqrt{2}}\sum_j q_j\cos(j\pi P_0(z)) \Big]. \label{eq:lqj}
\end{equation} The contribution to the Fisher matrix from this single
galaxy is
\begin{eqnarray}
F^{(1)}_{spec}(q_j,q_k) \! &=& \Big\langle
{\partial \ln{\cal L}\over \partial q_j} {\partial \ln{\cal L}\over
\partial q_k} \Big\rangle \nonumber \\ &=& \Big\langle \frac{\cos(j\pi
P_0(z))\cos(k\pi P_0(z))}{2} \Big\rangle = \frac{\delta_{jk}}{4}.\;\;
\end{eqnarray}
The last equality follows from the orthonormality of the
Fourier modes, combined with the fact that $P_0(z)$ is uniformly
distributed between $0$ and $1$.  If $N_{spec}$ galaxies have their
spectra measured, and these galaxies are drawn independently from the
source catalog, then the above Fisher matrix is multiplied by $N_{spec}$:
\begin{equation}
F^{(N_{spec})}_{spec}(q_j,q_k) = {1\over 4}N_{spec}\delta_{jk}. 
\end{equation}
In the
case of tomography, we have only $N_{spec}/2$ spectra in each of the two
tomography bins $A$ and $B$, and so:  
\begin{equation}
F^{(N_{spec})}_{spec}(q_j^A,q_k^A)
= {1\over 8}N_{spec}\delta_{jk},
\end{equation}
 and similarly for bin $B$.  In principle
it is possible to take different numbers of spectra in the two bins; we
have not attempted any optimization of this. We consider the cases of
$N_{spec}=0$, $64$, $512$ and $4096$ respectively.  The results can be compared to the
case where the source redshift distribution is known exactly by taking the
limit $N_{spec}\rightarrow\infty$.

The Fisher matrix is an asymptotic expansion, and while it is the standard tool of parameter forecasting, it sometimes leads to
over-optimistic parameter constraints (see Ref.~\cite{2003PhRvD..68h3002H} for an extreme example).  We have not tested its validity 
when used with this many parameters, although this could be done by Monte Carlo methods as used by Ref.~\cite{2004astro.ph..4317T}.

\section{Spectroscopy}
\label{sec:spectro}
In the discussion above, it has been assumed that the spectroscopically
obtained $n(z)$ is on average the same as the redshift distribution $n(z)$
of the sources used for lensing, and that the redshifts of the galaxies
targeted for spectroscopy are independent.  These assumptions will be
literally true in the idealized case that the spectroscopic galaxies are
chosen independently from the lensing source catalog, and there are no
failures to obtain spectro-$z$'s.  In practice, spectro-$z$s would
probably be obtained by a multi-object spectrograph attached to a large
telescope.  The spectro-$z$ failure rate will be minimized by using the
longest practical integration time to maximize signal-to-noise for each
spectrum, and the demands on telescope time are thus reduced if multiple
objects in the same field of view can be targeted simultaneously.  
However in this case the galaxies are not being drawn independently from
the source catalog, in particular large-scale clustering can cause the
redshifts of neighboring source galaxies to become correlated.  The
feasibility of obtaining many independent spectro-$z$'s is directly tied
to the clustering of the sources and the field of view of the telescope
since these determine the maximum number of galaxies that can be
simultaneously targeted for spectroscopy without the results being 
strongly correlated.

The effect of source clustering can be understood within the context of
Fisher matrix theory as follows.  If the fiducial model is correct, and 
the analysis is done assuming that the spectro-$z$s are independent, the 
difference $\delta p^\alpha = p^\alpha({\rm est}) - p^\alpha({\rm fid})$ 
between the estimated and fiducial model parameters is roughly
\begin{equation}
\delta p^\alpha \approx [{\bf F}^{-1}]^{\alpha\beta}
\left.{\partial\ln{\cal L}\over \partial p^\beta}\right|_{\rm fid},
\label{eq:dpa}
\end{equation}
where ${\cal L}$ is the likelihood assuming independent spectro-$z$s (i.e. 
the sum of Eq.~\ref{eq:lqj} over the spectroscopic targets), ${\bf F}$ is 
the Fisher matrix (also assuming independent spectro-$z$s), and the 
gradient is taken at the fiducial model.  So long as each galaxy in the 
lensing source catalog has an equal probability of being targeted for 
spectroscopy, it is easy to see that the spectroscopy contribution to 
$\langle {\partial\ln{\cal L}\over 
\partial q_j}\rangle = \frac{1}{\sqrt{2}}\sum_{a=1}^{N_{spec}}
\langle\cos(j\pi P_0(z_a))\rangle=0$ 
 and hence $\langle \delta p^\alpha 
\rangle = 0$, i.e. the inclusion of the spectroscopy likelihood function 
introduces no bias in the parameters even if the 
spectroscopic galaxies are not chosen independently.  However, the 
clustering of the sources does increase the covariance matrix of the 
$\delta p^\alpha$; taking the covariance of Eq.~(\ref{eq:dpa}) gives
\begin{eqnarray}
\langle \delta p^\alpha \delta p^\beta \rangle &\approx &
[{\bf F}^{-1}]^{\alpha\gamma}[{\bf F}^{-1}]^{\beta\delta}
\left\langle {\partial\ln{\cal L}\over \partial p^\gamma}
{\partial\ln{\cal L}\over \partial p^\delta} \right\rangle_{\rm fid}
\nonumber \\
&=&
[{\bf F}^{-1}]^{\alpha\gamma}[{\bf F}^{-1}]^{\beta\delta}
\Bigl\{ [{\bf F}_{CMB}+{\bf F}_{WL} + \Pi]_{\gamma\delta}
\nonumber \\ && + \delta_\gamma^{q_j}\delta_\delta^{q_k}
\sum_{a,b=1}^{N_{spec}} \frac{
\langle\cos(j\pi P_0(z_a))\cos(k\pi P_0(z_b))\rangle}{2}\Bigr\}.
\label{eq:pvar}
\end{eqnarray}
If we included only the $a=b$ terms in Eq.~(\ref{eq:pvar}), the last term 
in $\{\}$ would simply be the spectroscopy Fisher matrix for independent 
spectro-$z$s, $F_{spec,\gamma\delta}$.  The $a\neq b$ terms are only 
non-zero due to correlations of the galaxies with each other, and they 
contribute to quantity in $\{\}$ by an amount $\Delta_{q_jq_k}$.  Since 
the probability of two galaxies separated by angle $\theta_{ab}$ being 
physically associated with each other is 
$\omega(\theta_{ab})/[1+\omega(\theta_{ab})]$, where $\omega$ is the 
angular correlation function, and the cosine is bounded in the range $-1$ 
to $+1$, it follows that the $a\neq b$ contribution is
\begin{equation}
|\Delta_{q_jq_k}| \le {1\over 2}\sum_{a\neq b} \frac{\omega(\theta_{ab})}
{1 + \omega(\theta_{ab}) } \le {1\over 2}\sum_{a\neq b}\omega(\theta_{ab}),
\end{equation}
as compared with $F_{spec,q_jq_k}={1\over 4}N_{spec}\delta_{jk}$.  
Therefore if $\sum_{a\neq b}\omega(\theta_{ab})\ll N_{spec}/2$, then the 
clustering contribution $\Delta_{\gamma\delta}$ will be small compared to 
the spectroscopy Fisher matrix $F_{spec,\gamma\delta}$ and hence will 
contribute negligibly to the parameter uncertainties according to
Eq.~(\ref{eq:pvar}).

Let us consider the implications of this result for an idealized
spectroscopic survey strategy that consists of observing $J$ widely
separated fields, and obtaining $M=N_{spec}/J$ spectra in each field.  We
suppose the telescope has a circular field of view of angular radius $r$,
that the number density of galaxies targeted for selection in each field 
is $n=M/\pi r^2$, and that the angular correlation function is 
$\omega(\theta) = (\theta_0/\theta)^{0.7}$.  We then have
\begin{eqnarray}
\sum_{a\neq b}\omega(\theta_{ab})&\approx &
n^2J \int\int \left( \frac{\theta_0}{|\vec\theta_1-\vec\theta_2|} 
\right)^{0.7}\, d\vec\theta_1\, d\vec\theta_2
\nonumber \\ &=&
13.25\,n^2J\theta_0^{0.7}r^{3.3}
= 1.34\,{N_{spec}^2\over J}\left({\theta_0\over r}\right)^{0.7}.
\label{eq:wsum}
\end{eqnarray}
where the integrals are taken over a circular disk of radius $r$.  (The 
replacement of the galaxy-galaxy pair summation by an integral is 
appropriate because the integral is convergent at small 
separation $|\vec\theta_1-\vec\theta_2|$.)  The factor of $J$ in the first 
line occurs because we have to repeat the sum for each of the $J$ fields.  
In order to have $\sum_{a\neq b}\omega(\theta_{ab})\ll N_{spec}/2$, 
Eq.~(\ref{eq:wsum}) tells us that we need a number of fields given by
\begin{equation}
J\gg 2.68\,N_{spec}\left({\theta_0\over r}\right)^{0.7}.
\label{eq:jreq}
\end{equation}
Not surprisingly, the number of fields that need to be observed depends on
the ratio of the field-of-view radius $r$ to the angular clustering scale
$\theta_0$.  An angular clustering scale of $\theta_0\sim 0.0002$~degrees
is observed in SDSS \cite{2002ApJ...579...42C} for the magnitude range
$21<r<22$ (valid at separations of 1--30~arcmin).  Even smaller $\theta_0$
applies to fainter samples, e.g. the CFDF survey
\cite{2001A&A...376..756M} finds $\omega(1')\approx 0.01$ for
$18.5<I_{AB}<25$ galaxies, corresponding to $\theta_0=2.3\times
10^{-5}$~degrees (for slope $-0.7$, which is consistent with the
$\omega(\theta)$ data from $<1$ to several arcminutes).  The fainter CFDF sample is probably more representative of the galaxies that will be used in future WL surveys.
For $N_{spec}=512$ and a field of view of 0.5 degree radius, we find from
Eq.~(\ref{eq:jreq}) the requirement $J\gg 6$ ($\theta_0=0.0002$~deg) or
$J\gg 1.3$ ($\theta_0=2.3\times 10^{-5}$~deg).  In either case the number
of fields that must be observed to achieve $N_{spec}=512$ ranges from a
few to a few dozen, and for $N_{spec}=4096$ we find that the minimum number of fields $J$ is a few dozen to a few hundred.  The number of spectra to be obtained per field 
is $M=N_{spec}/J\le\sim 100$.

The problem of spectro-$z$ failures is more difficult to assess than the
source clustering.  In order to accurately reproduce the $n(z)$ of the
lensing source catalog, the targets in the associated spectroscopic survey
must be selected randomly from the lensing catalog, or at least have the
same selection criteria.  Spectro-$z$ failures are not part of the lensing
catalog selection criteria, and hence can bias the $n(z)$ determination.  
The spectro-$z$ failure rate can be reduced by using large-aperture
telescopes and very long integration times, which may prove feasible if 
$r$ is large, $\theta_0$ is small, and hence the number of fields $J$ to 
be observed is only a few.  It may also be possible to reduce failures by imposing appropriate color cuts on the source sample.
Nevertheless, the failure rate will never 
be exactly zero.  The treatment of systematic errors in $n(z)$ due to 
these failures is beyond the scope of this paper, but clearly deserves consideration in future work.

\section{Results}
\begin{table*}
{\scriptsize
\caption{\label{tab:notomo}Parameter estimation errors for WL+CMB: For WL, no-tomography, $f_{sky}=0.01,\;0.1$,
  and $\ell_{\rm max}=3000$. We use
  $N_{spec}=0$, 64, 512, 4096 and the limit $N_{spec}\rightarrow\infty$,
 and  a priors of 0.04 on $\cals$ and $\calr$ (unless indicated fixed). 
In the last part of the table, we show the results with the number of terms in the series 
increased from 5 to a 100.}
\begin{ruledtabular}
\begin{tabular}{ccccccccccccccccccc}
 & $f_{sky}$ &
$\Omega_m h^2$    &
$\Omega_b h^2$    & 
$\Omega_\Lambda$  &
$\sigma_8$        & 
$n_s$	          &
$\alpha_s$        &
$\tau$            &
$T/S$             &
$w$               &
$q_1$    &
$q_2$    &
$q_3$    &
$q_4$    &
$q_5$    &
$\sigma(\cals)$             &
$\sigma(\calr)$            \\
\hline
CMB only$\rightarrow$ &
&         0.0118& 0.0013& 0.1981&  0.086&  0.059&  0.036&  0.018&  0.187&    0.872&  -&  -&  - & - & - & - & - \\
\hline
$N_{spec}$=0&   0.01&    0.0096& 0.0009& 0.1320&  0.053&  0.033&  0.017&  0.018&  0.142&  0.514&  4.342& 19.912& 34.086& 66.902& 58.275&  0.040&  0.040\\
&   0.10&    0.0092& 0.0009& 0.0895&  0.039&  0.027&  0.012&  0.018&  0.129&  0.373&  1.595&  6.482& 13.804& 21.929& 18.556&  0.040&  0.040\\
\hline
$N_{spec}$=64&   0.01&    0.0083& 0.0008& 0.0248&  0.022&  0.022&  0.012&  0.017&  0.120&  0.226&  0.162&  0.244&  0.250&  0.249&  0.250&  0.040&  0.040\\
&   0.10&    0.0075& 0.0007& 0.0141&  0.018&  0.019&  0.010&  0.017&  0.114&  0.186&  0.121&  0.218&  0.249&  0.248&  0.250&  0.040&  0.039\\
\hline
$N_{spec}$=512&   0.01&    0.0076& 0.0008& 0.0219&  0.021&  0.021&  0.011&  0.016&  0.115&  0.220&  0.081&  0.088&  0.088&  0.088&  0.088&  0.040&  0.039\\
&   0.10&    0.0058& 0.0007& 0.0116&  0.014&  0.016&  0.010&  0.015&  0.098&  0.140&  0.074&  0.087&  0.088&  0.088&  0.088&  0.039&  0.038\\
\hline
$N_{spec}$=4096&   0.01&    0.0075& 0.0007& 0.0211&  0.021&  0.021&  0.011&  0.016&  0.114&  0.219&  0.031&  0.031&  0.031&  0.031&  0.031&  0.040&  0.039\\
&   0.10&    0.0049& 0.0006& 0.0111&  0.013&  0.014&  0.009&  0.015&  0.090&  0.118&  0.030&  0.031&  0.031&  0.031&  0.031&  0.039&  0.038\\
\hline
$N_{spec}\rightarrow\infty$&   0.01&    0.0074& 0.0007& 0.0210&  0.021&  0.020&  0.011&  0.016&  0.113&  0.219&  -    &  -    &  -    &  -    &  -    &  0.040&  0.039\\
&   0.10&    0.0047& 0.0006& 0.0110&  0.012&  0.013&  0.009&  0.015&  0.088&  0.113&  -    &  -    &  -    &  -    &  -    &  0.039&  0.038\\
\hline
$N_{spec}$=0&
   0.01&    0.0096& 0.0009& 0.1320&  0.053&  0.033&  0.017&  0.018&  0.142&  0.514&  4.341& 19.912& 34.086& 66.902& 58.275&  -    &  -    \\
$\calx\;$fixed&
   0.10&    0.0092& 0.0009& 0.0895&  0.039&  0.027&  0.012&  0.018&  0.129&  0.373&  1.594&  6.482& 13.804& 21.928& 18.555&  -    &  -    \\
\hline
$N_{spec}$=64&
   0.01&    0.0083& 0.0008& 0.0243&  0.022&  0.022&  0.012&  0.017&  0.120&  0.226&  0.159&  0.244&  0.250&  0.249&  0.250&  -    &  -    \\
$\calx\;$fixed&
   0.10&    0.0075& 0.0007& 0.0137&  0.018&  0.019&  0.010&  0.017&  0.114&  0.185&  0.114&  0.217&  0.249&  0.248&  0.250&  -    &  -    \\
\hline
$N_{spec}$=512&
   0.01&    0.0076& 0.0008& 0.0211&  0.021&  0.021&  0.011&  0.016&  0.115&  0.220&  0.081&  0.088&  0.088&  0.088&  0.088&  -    &  -    \\
$\calx\;$fixed&
   0.10&    0.0056& 0.0007& 0.0110&  0.014&  0.015&  0.009&  0.015&  0.096&  0.137&  0.073&  0.087&  0.088&  0.088&  0.088&  -    &  -    \\
\hline
$N_{spec}$=4096&
   0.01&    0.0074& 0.0007& 0.0202&  0.021&  0.020&  0.011&  0.016&  0.113&  0.219&  0.031&  0.031&  0.031&  0.031&  0.031&  -    &  -    \\
$\calx\;$fixed&   0.10&    0.0045& 0.0006& 0.0105&  0.012&  0.013&  0.009&  0.014&  0.085&  0.110&  0.030&  0.031&  0.031&  0.031&  0.031&  -    &  -    \\
\hline
$N_{spec}\rightarrow\infty$&    0.01&    0.0073& 0.0007& 0.0201&  0.021&  0.020&  0.011&  0.016&  0.113&  0.218&  -    &  -    &  -    &  -    &  -    &  -    &  -    \\
$\calx\;$fixed&   0.10&    0.0042& 0.0006& 0.0104&  0.012&  0.012&  0.009&  0.014&  0.083&  0.104&  -    &  -    &  -    &  -    &  -    &  -    &  -    \\
\hline
 & $f_{sky}$ &
$\Omega_m h^2$    &
$\Omega_b h^2$    & 
$\Omega_\Lambda$  &
$\sigma_8$        & 
$n_s$	          &
$\alpha_s$        &
$\tau$            &
$T/S$             &
$w$               &
$q_1$    &
$q_2$    &
to &
$q_{99}$    &
$q_{100}$    &
$\sigma(\cals)$             &
$\sigma(\calr)$            \\
\hline
$N_{spec}$=64&
   0.01&    0.0083& 0.0008& 0.0249&  0.022&  0.022&  0.012&  0.017&  0.120&  0.227&  0.162&  0.244& ...  &  0.250&  0.250&  0.040&  0.040\\
&   0.10&    0.0075& 0.0007& 0.0144&  0.018&  0.019&  0.010&  0.017&  0.114&  0.186&  0.122&  0.218&...  &  0.250&  0.250&  0.040&  0.039\\
\hline
$N_{spec}$=512&
   0.01&    0.0076& 0.0008& 0.0219&  0.021&  0.021&  0.011&  0.016&  0.115&  0.220&  0.081&  0.088& ... &  0.088&  0.088&  0.040&  0.039\\
&   0.10&    0.0058& 0.0007& 0.0116&  0.014&  0.016&  0.010&  0.015&  0.098&  0.140&  0.074&  0.087& ... &  0.088&  0.088&  0.039&  0.038\\
\hline
$N_{spec}$=4096&   0.01&    0.0075& 0.0007& 0.0211&  0.021&  0.021&  0.011&  0.016&  0.114&  0.219&  0.031&  0.031& ...  &  0.031&  0.031&  0.040&  0.039\\
&   0.10&    0.0049& 0.0006& 0.0111&  0.013&  0.014&  0.009&  0.015&  0.090&  0.118&  0.030&  0.031&  ... &  0.031&  0.031&  0.039&  0.038\\
\hline
$N_{spec}\rightarrow\infty$
&0.01&    0.0074& 0.0007& 0.0210&  0.021&  0.020&  0.011&  0.016&  0.113&  0.219&  - &  - &  ... &  - &  -  &  0.040&  0.039\\
&   0.10&    0.0047& 0.0006& 0.0110&  0.012&  0.013&  0.009&  0.015& 0.088&0.113&  - &  - &  ... &  - &  - &  0.039&  0.038\\
\end{tabular}
\end{ruledtabular}
}
\end{table*}
\begin{figure}[ht]
\includegraphics[width=2.4in,angle=-90]{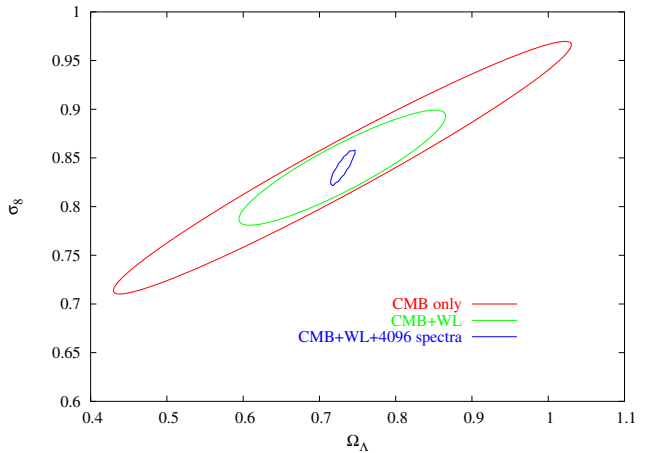}
 \caption{\label{fig:fig2} The 68.3\% confidence ellipses (assuming Fisher errors) for CMB and weak lensing.  Note the improvement in 
the constraints when spectra are available, versus the case where we marginalize over the redshift distribution parameters.
}
\end{figure}
\begin{figure}[ht]
\includegraphics[width=2.4in,angle=-90]{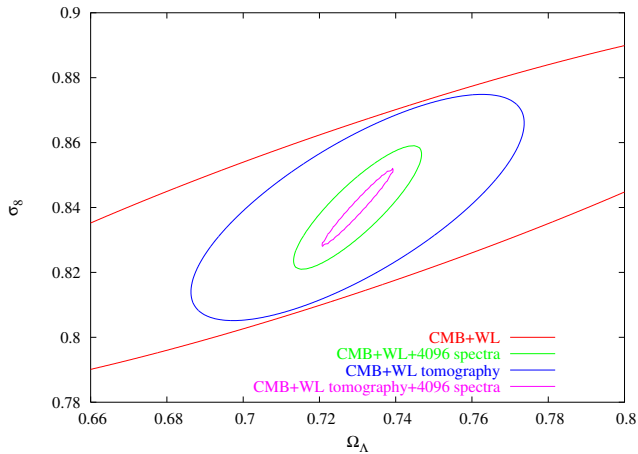}
 \caption{\label{fig:fig3} Same as Fig.~\ref{fig:fig2}, but including tomography.
}
\end{figure}
\begin{table}
{\scriptsize
\caption{\label{tab:tomo}Parameter estimation errors for WL+CMB: WL $f_{sky}$ is varied from
  0.01 to 0.1, and $\ell_{\rm max}=3000$. We use
  $N_{spec}=0$, 64, 512, 4096 and the limit $N_{spec}\rightarrow\infty$, and 
  priors of 0.04 on $\cals$ and $\calr$ (unless indicated fixed). 
  Tomgraphy case.}
\begin{ruledtabular}
\begin{tabular}{ccccc}
  &$f_{sky} $ &
$\Omega_\Lambda$  &
$\sigma_8$        & 
$w$               \\
\hline
CMB only$\rightarrow$&
& 0.1981&  0.086&   0.872\\
\hline
$N_{spec}=0$   
&   0.01&    0.0785&  0.035&  0.339\\
&   0.10&    0.0340&  0.024&  0.240\\

\hline
$N_{spec}=64$  
&   0.01&    0.0169&  0.018&  0.185\\
&   0.10&    0.0102&  0.013&  0.118\\

\hline
$N_{spec}=512$ 
&   0.01&    0.0160&  0.017&  0.173\\
&   0.10&    0.0084&  0.010&  0.088\\

\hline
$N_{spec}=4096$ 
&   0.01&    0.0154&  0.016&  0.162\\
&   0.10&    0.0078&  0.009&  0.074\\

\hline
$N_{spec}\rightarrow\infty$ 
&   0.01&    0.0148&  0.015&  0.154\\
&   0.10&    0.0071&  0.008&  0.066\\

\hline
\hline
$N_{spec}=0$   
&   0.01&    0.0684&  0.032&  0.311\\
$\calx\;$fixed&   0.10&    0.0288&  0.023&  0.230\\

\hline
$N_{spec}=64$
&   0.01&    0.0147&  0.017&  0.168\\
$\calx\;$fixed&   0.10&    0.0091&  0.012&  0.108\\
   
\hline
$N_{spec}=512$
&   0.01&    0.0137&  0.016&  0.153\\
$\calx\;$fixed&   0.10&    0.0069&  0.009&  0.075\\
                                                                                                                 
\hline
$N_{spec}=4096$
&   0.01&    0.0130&  0.014&  0.140\\
$\calx\;$fixed&   0.10&    0.0062&  0.008&  0.063\\
                                                                                                                 
\hline
$N_{spec}\rightarrow\infty$
&   0.01&    0.0118&  0.012&  0.122\\
$\calx\;$fixed&   0.10&    0.0043&  0.005&  0.043\\
   
\end{tabular}
\end{ruledtabular}
}
\end{table}
Our results are summarized in Tables~\ref{tab:notomo} and \ref{tab:tomo}, and 
Figs.~\ref{fig:fig2} and \ref{fig:fig3}. As expected, we find that 
combination of constraints from WL and from CMB leads to significant 
improvements in parameter estimation, notably for $\sigma_8$, $\Omega_m 
h^2$, $n_s$, $\alpha_s$, $w$ and $\Omega_{\Lambda}$.

As shown in Table \ref{tab:notomo}, the increase of the number of 
expansion terms from $j_{max}=5$ to 100 has little effect on this result.  
This suggests that the parameter estimates have converged and, since 
Eq.~(\ref{eq:z_dist_expansion}) can model an arbitrary function for 
sufficiently large $j_{max}$, it suggests that the form 
(Eq.~\ref{eq:z_dist_expansion}) for the redshift distribution is not 
artificially constraining the cosmological parameters.

\section{Discussion}

In agreement with previous results, we find that the addition of redshift information is helpful for several of the cosmological 
parameters such as $\Omega_{\Lambda}$, $\sigma_8$, $n_s$, and $w$. 
The optical depth $\tau$ is less improved by lensing because most of the statistical power on $\tau$ is coming from the CMB polarization 
reionization peak, which is not degenerate with any lensing-related quantities.  The baryon density is well-constrained by the CMB, 
however by improving our constraints on several cosmological parameters, lensing information breaks the relatively weak remaining 
degeneracies in the CMB and reduces the error from $\sigma(\Omega_bh^2)=0.0013$ to $0.0009$.  Even lensing with $N_{spec}=0$ is 
sufficient to break this degeneracy, so inclusion of redshift information adds little for $\Omega_bh^2$.

The addition of redshift information leads to further significant 
improvements in the parameter estimation as expected.  The $N_{spec}=0$ 
constraints are substantially worse than $N_{spec}=64$, because in the 
$N_{spec}=0$ case even wildly oscillating redshift distributions $n(z)$ 
are formally allowed, as is evidenced by the large $\sigma(q_j)\gg 1$ in 
Table~\ref{tab:notomo}.  Although the gradual addition of spectroscopic 
redshifts does make additional improvements on the constraints on the 
cosmological parameters, we find that there is a certain number of 
$N_{spec}$ (several thousands for the surveys considered here) 
beyond which the addition of further spectra will make only a very small 
improvement to the cosmological parameters.

In order to try to explain this, let us recall that  
error bars are correlated, and hence there can still be 
combinations of parameters that are degraded by incomplete knowledge of
the source redshift distribution.  Information on this degradation is
contained within the $9\times 9$ covariance matrix $\textbf{C}$ of the
cosmological parameters $\{\Omega_mh^2, \Omega_bh^2, \Omega_\Lambda,
\sigma_8, n_s, \alpha_s, \tau, T/S, w\}$.
 We examine the degradation factor $R$ 
of the combination of parameters $x=k_\alpha p^\alpha$ given by
\begin{equation}
R(x) = \frac{\sigma^2(x;N_{spec})}{\sigma^2(x,N_{spec}=\infty)}
= \frac{[\textbf{C}(N_{spec})]^{\alpha\beta}k_\alpha k_\beta}{
[\textbf{C}(\infty)]^{\alpha\beta}k_\alpha k_\beta}.
\end{equation}
The maximum value of $R$ can be found by setting $\partial R/\partial 
k_\alpha=0$, which leads to the eigenvalue equation
\begin{equation}
\frac{[\textbf{C}(N_{spec})]^{\alpha\beta}k_\alpha k_\beta}{
[\textbf{C}(\infty)]^{\alpha\beta}k_\alpha k_\beta}k_\gamma
= [\textbf{C}^{-1} (\infty)]_{\gamma\beta}
[\textbf{C}(N_{spec})  ]^{\beta\alpha}k_\alpha.
\end{equation}
Thus the $k_\alpha$ that maximizes the degradation $R(x)$ is an 
eigenvector of $[\textbf{C}^{-1}(\infty)\textbf{C}(N_{spec})]$, and the 
degradation factor $R(x)$ is the eigenvalue.  (The maximum degradation 
factor corresponds to the maximum eigenvalue of 
$[\textbf{C}^{-1}(\infty))\textbf{C}(N_{spec})]$; the other eigenvectors 
are minima or saddle points of $R$.)  For $N_{spec}=512$, $f_{sky}=0.1$, 
and no-tomography,
we find a maximum eigenvalue of $\lambda_{max}=2.2$ with the corresponding 
combination of parameters
\begin{eqnarray}
x &=& -4.37\Omega_mh^2+8.89\Omega_\Lambda-5.39\sigma_8 - 2.57n_s
+4.32\alpha_s
\nonumber \\ &&
+0.53w + 3.11\Omega_bh^2 - 4.59\tau-7.18T/S.
\end{eqnarray}
Having only a finite number of spectra is degrading the 
$1\sigma$ WL+CMB constraint on $x$ by a factor of 
$\sqrt{\lambda_{max}}\approx 1.5$, but from Table~\ref{tab:notomo} we can 
see that the 
constraints on the standard set of cosmological parameters is degraded by 
$<10\%$.  This is because $x$ is a direction in parameter space that is 
very well-constrained by WL+CMB: with CMB only, we have 
$\sigma(x)=1.02$, whereas with WL and $N_{spec}=512$ added to 
the CMB we have 
$\sigma(x)=0.098$ and correlation coefficient $\rho(x,q_1)=-0.64$.  
Consequently, although $x$ is degraded by imperfect knowledge of 
the source redshift distribution, and is degenerate with the redshift 
distribution parameters, it is not the direction that dominates 
the uncertainties on the individual cosmological parameters. 
The second-largest eigenvalue of 
$[\textbf{C}^{-1}(\infty))\textbf{C}(N_{spec})]$ is 1.2, indicating that 
the directions other than $x$ are not significantly degraded.
When we fix the calibration 
parameters, the eigenvector analysis shows an increase of almost
an order of magnitude on the degradation with 
$\sqrt{\lambda_{max}}\approx 16.6$, but again, not in the direction 
that dominates the uncertainties on the conventional cosmological 
parameters.  As seen from Table \ref{tab:notomo} and
Table \ref{tab:tomo}, a better calibration will make a 
larger numbers of spectra slightly more useful but a significant 
improvement require a rather idealized perfect knowledge of 
the calibration.

The most significant improvement in the cosmological parameters with large numbers of spectra comes from the cases with
tomography, since this helps break the degeneracies that dominate the parameter uncertainties and allows the directions well
constrained by WL to become more important.  One can see from Table~\ref{tab:tomo} that the uncertainties on $\Omega_\Lambda$, 
$\sigma_8$, and $w$ are degraded by $\sim 50\%$ for $N_{spec}=4096$ versus $N_{spec}=\infty$ with the larger area $f_{sky}=0.1$.  In all 
the other cases the degradation with $N_{spec}=4096$ is much less.

It is possible that different results would be obtained if WL were combined with additional cosmological probes such as supernovae, 
which could break remaining degeneracies in the data and therefore make the degradation of $x$ more important.  A general way to 
investigate this possibility is to examine $\lambda_{max}$ for the WL-only matrices, since combining WL with cosmological probes that 
do not include the $\{q_j\}$ must result in a degradation factor $\lambda_{max}$ smaller than that for WL alone. When we apply the 
eigenvalue analysis to the weak lensing only matrices, we obtain the degradation factors of $\sqrt{\lambda_{max}}\approx 19.98$ for 
the fixed calibration case and $N_{spec}=512$.  The latter number indicates that there are directions in cosmological parameter space 
which are dramatically improved by exact knowledge of the redshift distribution rather than only 512 spectra.  We can reduce the 
degradation factor to $\lambda_{max}=2$ by increasing $N_{spec}$ to 200000, and it is thus only for $N_{spec}\ge 200000$ that {\em every} 
direction in parameter space is limited by lensing statistics rather than redshift distribution uncertainties.  However, this 
improvement is rapidly lost due to the approximate calibration-redshift degeneracy if the calibration is uncertain: if the prior on 
the calibration parameters is widened to $\sigma(\zeta_s)=\sigma(\zeta_r)=0.02$ we achieve $\lambda_{max}=2$ at $N_{spec}=4000$, and 
with $\sigma(\zeta_s)=\sigma(\zeta_r)=0.04$ we achieve $\lambda_{max}=2$ at $N_{spec}=1100$.  Thus having either good calibration or a 
well-determined redshift distribution individually may not be very useful, but having both combined can significantly improve some 
constraints.

We conclude that, though significant improvement is obtained from the addition of redshifts information, there is a certain number of 
$N_{spec}$ -- of order $10^3$ for the surveys considered here -- beyond which the addition of further spectra will make only a very 
small improvement to the cosmological parameters.  We do find that there are directions in parameter space that continue to improve and 
do not saturate until $N_{spec}$ is very large, especially if the calibration is very well determined; however these directions 
correspond to very large eigenvalues of the WL Fisher matrix and do not dominate the parameter uncertainties in the WL+CMB 
combinations we have discussed.  The results presented here indicate that if $N_{spec}\sim$few$\times 10^3$ spectra of representative 
sources can be obtained, cosmological constraints can potentially be improved relative to the CMB-only case without model-dependent 
assumptions about the source redshift distribution.  These results are robust against fluctuations in the spectro-$z$ distribution due to galaxy clustering so long as 
enough spectroscopic fields are observed, as detailed in Sec.~\ref{sec:spectro}, while future work is required 
to address the spectro-$z$ failures.

\acknowledgments

We thank Uro\v{s} Seljak and David Spergel for useful comments.
M.I. acknowledges the support of the Natural Sciences and Engineering
Research Council of Canada (NSERC). C.H. acknowledges the support of the
National Aeronautics and Space Administration (NASA) Graduate Student
Researchers Program (GSRP).

\end{document}